\documentstyle[twocolumn,prc,aps,epsfig]{revtex}
\input epsf
\newcommand{\be}{\begin{eqnarray}}
\newcommand{\ee}{\end{eqnarray}}
\def\lsim{\mathrel{\rlap{\lower3pt\hbox{\hskip1pt$\sim$}}
     \raise1pt\hbox{$<$}}} 
\def\gsim{\mathrel{\rlap{\lower3pt\hbox{\hskip1pt$\sim$}}
     \raise1pt\hbox{$>$}}} 

\begin{document}

\twocolumn[\hsize\textwidth\columnwidth\hsize\csname @twocolumnfalse\endcsname

\title{Towards the Theory of Binary Bound States in Quark-Gluon Plasma} 

\author {Edward V.Shuryak and Ismail Zahed}
\address { Department of Physics and Astronomy\\ State University of New York,
     Stony Brook, NY 11794-3800}

\date{\today}
\maketitle
\begin{abstract}

Although at temperatures $T\gg \Lambda_{QCD}$ the quark-gluon plasma (QGP) 
is a gas of weakly interacting quasiparticles (modulo long-range magnetism), 
it is strongly interacting in the regime $T=(1-3)\,T_c$. As both heavy ion
experiments and lattice simulations are now showing, in this region the QGP 
displays rather strong interactions between the constituents. 
In this paper we investigate the relationship between four 
(previously disconnected) lattice results: 
{\bf i.} spectral densities from MEM analysis of correlators;
{\bf ii.} static quark free energies $F(R)$; 
{\bf iii.}  quasiparticle masses;
{\bf iv.}  bulk thermodynamics  $p(T)$.
We show a high degree of consistency among  them not known before.
The potentials $V(R)$ derived from $F(R)$ lead
to large number of binary bound states, mostly colored, in
$gq, qq, gg$, on top of the usual $\bar q q$ mesons.
Using the Klein-Gordon equation and ({\bf ii-iii})
we evaluate their binding energies and 
locate the zero binding endpoints on the phase diagram, which
happen to agree with ({\bf i}). We then
estimate the contribution of all states to the bulk thermodynamics
in agreement with ({\bf iv}).  
We also address a number of theoreticall issues related with to the
role of the quark/gluon spin in binding at large $\alpha_s$, although
we do not yet include those in our estimates. Also
the issue of the transport properties (viscosity, color conductivity) in 
this novel description of the QGP  will be addressed elsewhere.
\end{abstract}
\vspace{0.1in}
]
\begin{narrowtext}
\newpage
\section{Introduction}
\subsection{New view of the Quark-gluon Plasma at not-too-high T}

From its inception two decades ago, the high-T phase of QCD commonly
known as the Quark-Gluon Plasma (QGP) after~\cite{Shu_QGP}, 
was described as a weakly interacting gas of ``quasiparticles'' 
(quarks and gluons). Indeed, at very high temperatures asymptotic
freedom causes the electric coupling to be small and the QGP to
be weakly interacting or perturbative~\footnote{The exception 
is long range color magnetism which remains
nonperturbative~\cite{LINDE}.}. At intermediate temperatures
of few times the critical temperature $T_c$ of immediate relevance 
to current experiments, there is new and growing evidence that 
the QGP is not weakly coupled. In a recent letter~\cite{IS_newqgp} (to be
referred to below as SZ1) we have proposed that in this region QCD 
seems to be  close to {\em a strongly coupled Coulomb regime}, with 
an effective coupling constant $\alpha\approx 0.5-1$  
and multiple {\em bound states} of quasiparticles. We have argued there
(and will show it more quantitative below) that these bound states
are very important for the thermodynamics of the QGP,
responsible for roughly half of the pressure in RHIC domain. We will
not address transport properties of the QGP in this paper, although
we expect the bound states to play an even more important role.

To set the stage we first recall the chief arguments behind a 
scenario of {\it strongly coupled QGP} at such temperatures:
 {\bf i.} a low viscosity argument;
{\bf ii.} recent lattice findings of $\bar c c,\bar q q$  
bound states at $T>T_c$; {\bf iii.} the high pressure puzzle.

{\bf Transport properties} of QGP were so far studied mostly perturbatively,
in powers of the weak coupling. This approach predicted
a large mean free path,  $T\,l_{\rm mfp}\approx
1/g^4 {\rm ln}\,(1/g) \gg 1$. Similar pQCD-inspired ideas have
led to the pessimistic expectation that the Relativistic 
Heavy Ion Collider (RHIC) project in Brookhaven National Laboratory, 
would produce a firework of multiple  mini-jets rather than QGP. 
However already the very first RHIC  run, in the summer of 2000, has shown
spectacular collective phenomena known as
radial and elliptic flows. The spectra of about 99\%
of all kinds of secondaries (except their high-$p_t$ tails)
are accurately described by {\em ideal hydrodynamics}~\cite{hydro}.
Further studies of partonic cascades~\cite{GM} 
and viscosity corrections~\cite{Teaney_visc} 
have confirmed that one can only understand RHIC data by very low viscosity 
or large parton rescattering cross sections exceeding
pQCD predictions by large factors of about $\sim 50$ or so.
In short the QGP probed at RHIC is by far the {\em perfect liquid} known
so far, with the smallest viscosity-to-entropy ratio ever,
i.e.  $\eta/s\approx .1$~\cite{Teaney_visc}. 
We note that from the theory standpoint ideal hydrodynamics,
complemented by a ``non-ideal'' expansion in powers of 
the mean free path (the inverse powers of the rescattering cross section), is 
perhaps the oldest example of a {\em strong coupling} expansion.

Naturally, these observations have increased our interest in
other strongly interacting systems. Two such examples,
discussed already in SZ1 are: {\bf i.} trapped ultracold atoms 
driven to strong coupling via Feshbach resonances\cite{ATOM1,ATOM2}; {\bf ii.}
$\cal N$=4 supersymmetric gauge theory (CFT) recently studied 
via AdS/CFT correspondence~\cite{ADSCFT1,ADSCFT2}. In both cases,
see the atomic experiments~\cite{atoms_flow} for the first 
and the CFT viscosity calculation in~\cite{PSS} for the second,
strong coupling was found to lead to a hydrodynamic behavior,
with a very small viscosity.

The central point  of the SZ1 paper was to provide at least a qualitative
explanation to this small viscosity by relating it to 
multiple {\em loosely bound} binary 
states of quasiparticles, which should result
in larger scattering lengths 
induced by low-lying resonances. We argued that 
at the {\em zero binding points}
(indicated by lines in  Fig.~\ref{fig_masses_T}a and
black dots in  Fig.~\ref{fig_masses_T}b) those effects
should be enhanced by a
Breit-Wigner resonance, producing the cross section
 (modulo the obvious spin factors
depending on the channel) 
\be 
\sigma(k)\approx  {4\pi \over k^2} {\Gamma_i^2/4 \over (E-E_r)^2+\Gamma_t^2/4}
\ee
At  $E-E_r\approx 0$ the in- and total
widths are about equal and approximately cancel. The ensuing  
``unitarity limited'' scattering 
cross sections are large. This conjecture is nicely supported by the atomic experiments 
mentioned in~\cite{atoms_flow}, in which precisely the same mechanism
was shown to ensure a hydrodynamical ``elliptic flow''.

\begin{figure}[t]
\centering
\includegraphics[width=6.cm]{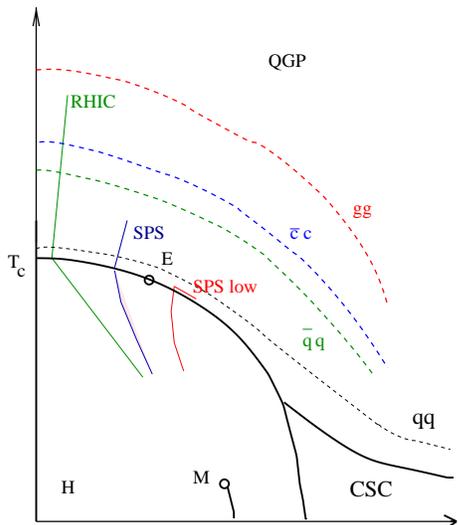}
   \caption{\label{fig_masses_T}
Schematic position of several  zero binding lines on the 
QCD phase diagram, from SZ1.
}
\end{figure}

 {\bf Bound states
in the QGP phase}. The earliest QGP signal (suggested by one of us
\cite{Shu_QGP}) was the {\em disappearance} of familiar hadronic states, 
especially the vector ones  -- $\rho,\omega,\phi,J/\psi$ mesons -- directly observable via
dilepton experiments. It was suggested later
that even  small-size and deeply-bound states of charmonium
such as $\eta_c,J/\psi$ were expected to melt at $T\approx T_c$
\cite{MS,KMS},
so their absence was proposed to be a ``golden signature'' of the QGP.   
However, recent lattice works~\cite{charmonium}
using the Maximal Entropy Method (MEM) have found that 
charmonium states actually persists to at least $T\approx 2\,T_c$, and there
are similar evidences about mesonic bound states made of light quarks as
well~\cite{Karsch:2002wv}. As we will show below,
this  is in good agreement
with independent lattice studies on the effective interaction potential
between static sources in QGP~\footnote{Discrepancies
with earlier results are mainly due to a confusion between
a free and potential energy.}.
In  SZ1 and now we argue that on top of familiar
colorless hadronic states there should
also be litterally hundreds of colored binary bound states, and  for the 
singlet $gg$ pair (the most attractive channel) the region of binding should persist 
up to quite high temperature, i.e. $T\approx 4\, T_c$.

\begin{figure}[t]
\centering
\includegraphics[width=7.cm]{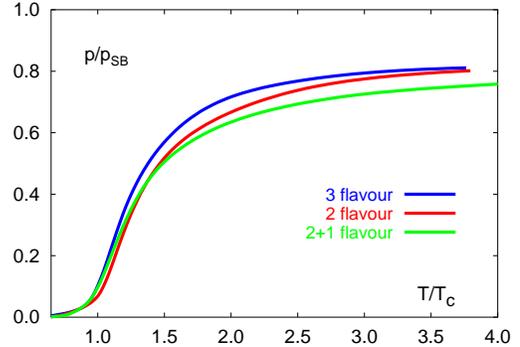} 
  \caption{\label{fig_lat_pressure}
The pressure normalized to that of  a gas of massless and
noninteracting quasiparticles, or Stephan-Boltzmann value, versus
the temperature $T/T_c$, from the lattice calculations
by the Bielefeld group~\protect\cite{Karsch_pressure}. 
Different curves are for different 
numbers (masses) of the dynamical quarks.
}
\end{figure}

 {\bf The ``high pressure puzzle''} stems from 
Fig.~\ref{fig_lat_pressure} which shows a sample of lattice results
for the QCD bulk pressure $p(T)$, normalized to Stephan-Boltzmann
noninteracting value for each theory. At 
$T\approx 2\,T_c$ the pressure nearly saturates to 0.8 of 
 that for massless noninteracting gas, $p_{SB}$. 

This behaviour appears to be in contradiction  with other
lattice QCD works which found rather heavy quasiparticles, with masses
(energies) e.g. of $M_{q,g}/T\sim 3$ at $T\approx 1.5-2\,
T_c$~\cite{Karsch_quasiparticles}. How could such  heavy quasiparticles
account for large 
pressures~\cite{Karsch_pressure}? Substantial elliptic flow effects at RHIC 
\cite{hydro} points also to a large pressure in the prompt phase
at RHIC or at $T=(1-2)\, T_c$, so we do know it is there.

A similar  discrepancy  but now analytic and parametric,
was found for CFT at parametrically large coupling.  
In our second paper~\cite{SZ_CFT} we argued that in this case
the matter cannot be made of quasiparticles, which are again too heavy
{\em parametrically},
but rather by their (much lighter) binary composites.
The QGP picture we advocate is thus just a beginning of this 
parametric trend, with
the intermediate running gauge coupling $\alpha_s\approx (1/2-1)$.

\section{Color forces in various binary channels}

At $T>T_c$ there is no color confinement, and so the 
interaction is a Coulomb-like at small distances, with a Debye-type screening
\cite{Shu_JETP} at large distances. Both lattice data as we will use
below and results from the AdS/CFT correspondence agree that these features of the
potential carry to the strong coupling regime.

In this section we compare multiple binary colored
channels~\footnote{Multi-body bound states are also
allowed, but will not be discussed here.}, by using 
{\em Casimir scaling} for their relative strength.
Its precise formulation can be made as follows.
Let $A,B$ be the color representations of either
the quark or gluon constituents. If they are
 in a colored bound state with
 overall color representation $D$, then their color interaction is
 proportional
to~\footnote{This formula is analogous to the familiar SU(2) 
calculation of a relative spin  projections
in a state with some total spin $J$.}
 
\be
{\bf c}_D=\left(\vec{\lambda}_A\cdot\vec{\lambda}_B \right)_D =
2\left({\bf C}(D) -{\bf C}( A) -{\bf C} (B)\right)\,\,\,,
\label{cb1}
\ee
where the ${\bf C}$'s are the pertinent Casimirs. For $SU(3)_c$ the
Casimirs can be  given in terms of the Dynkin index $(mn)$ of
the representation $D$, 

\be
{\bf C}(D) = m+n+\frac 13 (m^2+n^2+mn) 
\label{cb0}
\ee
In this section, we detail the color representations and the strength 
of the Coulomb interaction (\ref{cb1} in the bound states $\bf {gg}$, $\bf {qg}$
and $\bf {qq}$.

\vskip .5cm
{$\bf gg\,\,\,:$} Two gluons yield the sum of irreducible color representations
$D={\bf 8}\otimes {\bf 8}= {\bf 1}\oplus {\bf 8}_S \oplus {\bf 8}_A
\oplus {\bf 10}_S \oplus \overline{\bf 10}_A \oplus {\bf 27}$. In terms
of the Dynkin indices, the same irreducible decomposition yields
$D=(00) \oplus (11)\oplus (11) \oplus (03) \oplus (30) \oplus (22)$.
Thus (\ref{cb1}) reduces to

\be
{\bf c}_D=\left(\vec{\lambda}_8\cdot\vec{\lambda}_8 \right)_D =
2\left({\bf C}(D) -2\,{\bf C}(8)\right)\,\,\,,
\label{cb2}
\ee
with ${\bf c}_1=-12$, ${\bf c}_8=-6$ (attractive) and
${\bf c}_{10}=0$ (inactive) and ${\bf c}_{27}=4$ (repulsive).

\vskip .5cm
{$\bf qg\,\,\,:$} A quark and a gluon 
yield the sum of irreducible color representations
$D={\bf 3}\otimes {\bf 8}= {\bf 3}\oplus {\bf 6}_S \oplus {\bf 15}$. 
In terms of the Dynkin indices, the same irreducible decomposition yields
$D=(10) \oplus (02) \oplus (21)$. Thus (\ref{cb1}) reduces to

\be
{\bf c}_D=\left(\vec{\lambda}_3\cdot\vec{\lambda}_8 \right)_D =
2\left({\bf C}(D) -4/3-3\right)\,\,\,,
\label{cb22}
\ee
with ${\bf c}_3=-6$, ${\bf c}_6=-2$ (attractive) and
${\bf c}_{15}=2$ (repulsive). A similar decomposition applies
to the conjugate representation $\overline{\bf q}{\bf g}$ with
$D=(01) \oplus (20) \oplus (12)$

\vskip .5cm
{$\bf \overline{q}q\,\,\,:$} A quark and an antiquark  
yield the sum of irreducible color representations
$D=\overline{\bf 3}\otimes {\bf 3}= {\bf 1}\oplus {\bf 8}$. 
In terms of the Dynkin indices, the same irreducible decomposition yields
$D=(00) \oplus (11)$. Thus (\ref{cb1}) reduces to

\be
{\bf c}_D=\left(\vec{\lambda}_{\overline 3}\cdot\vec{\lambda}_3 \right)_D =
2\left({\bf C}(D) -8/3\right)\,\,\,,
\ee
with ${\bf c}_1=-16/3$ (attractive) and  ${\bf c}_8=+2/3$ (repulsive).

\vskip .5cm
{$\bf {q}q$\,\,\,.} A two-quark  state 
yield the sum of irreducible color representations
$D={\bf 3}\otimes {\bf 3}= \overline{\bf 3}\oplus {\bf 6}$. 
In terms of the Dynkin indices, the same irreducible decomposition yields
$D=(01) \oplus (20)$. Thus (\ref{cb1}) reduces to

\be
{\bf c}_D=\left(\vec{\lambda}_{3}\cdot\vec{\lambda}_3 \right)_D =
2\left({\bf C}(D) -8/3\right)\,\,\,,
\ee
with ${\bf c}_{\overline{3}}=-8/3$ (attractive) and  ${\bf c}_6=+4/3$ (repulsive).

Using a singlet $\bar{q} q$ as a standard benchmark (the only one studied 
extensively on the lattice), one can summarise
the list of all attractive channels in 
Table~\ref{tab_channels}, indicating the relative strength
of the Coulomb potential in a given color channel and the number of states.
Even without the excited states to be discussed below,
there is a total of 481 channels for two flavors (41 colorless),
and 749 states (81 colorless) for three flavors.

\section{Bound states in strong Coulomb field }

Before we analyze the relativistic two-body bound states for
quarks and gluons, we go over the results for the simpler problems 
involving either a spin 0, spin 1/2 and  spin 1 particle moving
in a strong (color) Coulomb field, where the effect of color
is treated as a Casimir rescaling of the Coulomb charge. 
Precession in color space can be treated but will be ignored
through an ``Abelianization'' of the external field. The 
results for spin 0 and 1/2 are known from 1928~\cite{solved_Dirac}.
They are presented for completeness since they streamline
our analysis for spin 1. A canonical application of the latter
is that of a W boson bound to heavy Coulomb center.

\begin{table}[t]
\begin{tabular}{llll}
channel & rep. & charge factor & no. of states \\  \hline
$gg$ &  1 & 9/4 & $9_s$ \\
$ gg$ &  8 & 9/8 & $9_s*16$                \\  \hline
$qg+\bar q g$ &  3 & 9/8 & $3_c*6_s*2*N_f$ \\
$qg+\bar q g$ &  6 & 3/8 & $6_c*6_s*2*N_f$        \\  \hline
$\bar q q$ &  1 & 1 & $8_s*N_f^2$  \\  \hline
$qq+\bar q \bar q$  &  3 & 1/2 & $4_s*3_c*2*N_f^2$  \\
\end{tabular}
\caption{Binary attractive channels discussed in this work, the subscripts
s,c,f mean spin,color and flavor, $N_f=3$ is the number of 
relevant flavors. 
\label{tab_channels}}
\end{table}


\subsection{Spin 0}

For a scalar particle the Klein-Gordon equation  
\be 
\left((E-gV)^2-m^2+\vec {\vec{\partial}}^2\right)\,\Phi=0 
\ee
should be used. This equation was analyzed for a Coulomb
potential\footnote{
We use in this section
a QED-like notations in which $\alpha=g^2$. The standard
QCD notations for fundamental charges should read $\alpha=(4/3)g^2/4\pi=(4/3)\alpha_s$. 
} 
$gV=-\alpha/r$ in~\cite{solved_Dirac} with the energy spectrum
\footnote{In~\cite{SZ_CFT} a WKB analysis was used with apologies
to~\cite{solved_Dirac}.}

\be 
{E(n_r,l)\over m}\, 
\left(1+{\alpha^2 \over (n_r+\sqrt{(l+1/2)^2-\alpha^2})^2}
\right)^{1/2}=1
\ee
Taking the lowest level to be
$n_r=1,l=0$ as an example, one finds that $\alpha=1/2$ is a critical
value for this equation.  Although the binding is  at this point
finite and even not large, $E(1,0)/m=\sqrt{4/5}$, something 
new is obviously happening at this
critical coupling  because the square root (in the denominator) goes
complex. 

What happens is that the particle starts  falling towards the center.
Indeed, ignoring at small $r$ all terms except the $V^2$ term one finds
that the radial equation is

\be 
R''+{2\over r} R'+{\alpha^2\over r^2}R=0
\ee
which at small $r$ has a general solution 

\be 
R=A r^{s_+}+B r^{s_-},\hspace{.5cm} s_\pm=-1/2\pm\sqrt{1/4-\alpha^2}
\ee
that for $\alpha\rightarrow 1/2$ is just $1/r^{1/2}$.
At the critical coupling  $both$ solutions have the same (singular) 
behavior at small $r$. For $\alpha > 1/2$ the falling starts, as one
sees from the complex (oscillating) solutions.

In the CFT theory with fixed (non-running)  coupling constant, nothing 
can prevent the particle from falling to arbitrary small r as soon as 
$\alpha> 1/2$~\footnote{In fact, this is why
the dual string description has a black hole. One of us (ES) thanks
Daniel Kabat for pointing this  to him.}.
In contrast, in QCD the coupling runs, $\alpha(r)\approx
1/{\rm ln}(1/r\Lambda_{QCD})$, so that at small enough distances 
the coupling gets less than critical and the falling stops.
In~\cite{BLRS} a crude model of a regularized Coulomb field was used,
producing the same effect. Our arguments show that asymptotic freedom 
would be in principle enough. However, the wave function at the origin
is changing dramatically at $\alpha\approx 1/2$ and in view of  that
we performed an additional study of the Klein-Gordon problem with a 
Coulomb+quasi-local potential in the Appendix. 

The falling onto the center happens for any spin of the particle, only
the value for the critical coupling is different.  We now proceed to
show that.

\subsection{Spin $\frac 12$}

The squared Dirac equation for a massive spinor ${\bf \Psi}$ reads 

\be
\left(\Box +m^2-\frac g2\,\sigma_{\mu\nu}\,F^{\mu\nu}\right) \,{\bf
\Psi} =0\,\,,
\label{D1}
\ee
with $\sigma_{\mu\nu}=i[\gamma_\mu,\gamma_\nu]/2$ and $F^{\mu\nu}$ 
the external background field. In the chiral basis, (\ref{D1}) 
simplifies

\be
\left((\Box+m^2) + 2g\,\vec{\bf S}\cdot (\vec{\bf B}\mp i\vec{\bf E})\right)
{\bf \Psi}_\pm =0\,\,,
\label{D2}
\ee
where the spin operator is Lie-algebra valued,

\be
\left[{\bf S}^a,{\bf S}^b\right] =if^{abc}\,{\bf S}^c
\label{D3}
\ee
The magnetic
contribution in (\ref{D1}) is standard. The electric contribution
is complex~\footnote{Recall that the squared Dirac operator is
hermitean.} and is reminiscent of a Bohm-Aharanov effect.

The relativistic 
Coulomb problem stemming from (\ref{D1}) for each of the two
stationary spin components reads

\be
\left((E+\frac{g^2}{r})^2-m^2 +\frac {d^2}{dr^2} + \frac 2r 
\frac {d}{dr} -\frac{{\vec{\bf L}}^2}{r^2} -\frac{2ig^2}{r^2}
\vec{\bf S}\cdot\hat{r}\right){\bf \Psi} =0\,\,.
\nonumber\\
\label{D4}
\ee
the solutions to (\ref{D4}) are naturally sought in terms of
spinor spherical harmonics~\cite{Edmonds}

\be
{\bf \Psi}^a_{JML} = \sum_{b}
\left(\matrix{J&L&\frac 12\cr
-M&(M-b)&b\cr}\right)\,{\bf e}_b^aY_L^{M-b}\,\,,
\label{D5}
\ee
with the bracket a conventional Clebsch-Gordon coefficient
restricting the values of $L$
and $a,b=\pm 1/2$. In this representation the spinors are just
${\bf e}^a_b=\delta^{ab}$, and (\ref{D5}) is an eigenstate of 
${\bf J}^2$, ${\bf J}_3$ and ${\bf L}^2$ with eigenvalues
$J(J+1)$, $-M$ and $L(L+1)$ respectively. In the basis (\ref{D5})
the spin operator ${\bf S}\cdot \hat{r}$ is off-diagonal

\be
{\bf S}\cdot\hat{r}= \left(\matrix{0&-\frac 12\cr 
-\frac 12&0\cr}\right)\,\,.
\label{D6}
\ee

Using the spinor spherical harmonics (\ref{D5}) in (\ref{D4})
through the expansion

\be
{\bf {\Psi}} =\sum_{a}\,{\bf \Psi}^a_{JML}\,{\bf R}^a 
\label{D7}
\ee
for fixed $JM$ and (\ref{D6}) we obtain the $2\times 2$ matrix
equation for the radial function ${\bf R}$

\be
\left((E+\frac{g^2}{r})^2 -m^2+\frac {d^2}{dr^2} + \frac 2r 
\frac {d}{dr} -\frac{J(J+1) +{\bf C}}{r^2} \right) \,{\bf R}=0
\nonumber\\
\label{D8}
\ee
with

\be
{\bf C} = \left(\matrix{\frac 14 -(J+\frac 12)&-ig^2\cr
-ig^2&\frac 14 + (J+\frac 12)\cr}\right)\,\,.
\label{D9}
\ee
The eigenvalues of ${\bf C}$ are

\be
\lambda_{\pm\frac 12} =\frac 14 \pm \sqrt{(J+\frac 12)^2-g^4}
\label{D10}\,\,.
\ee

In terms of (\ref{D10}) the eigenvalue equation (\ref{D8}) 
becomes diagonal for the rotated ${\bf R}$. Defining 
$\epsilon=\sqrt{m^2-E^2}$ and $r=x/2\epsilon$ and 
${\bf R}=2\epsilon\,{\bf U}/\sqrt{x}$ yield (\ref{D8})
in the diagonal basis in the form

\be
x{\bf U}''+ {\bf U}'+\left(\alpha-\frac
x4-\frac{\beta^2}{4x}\right)\,{\bf U} =0
\label{D11}
\ee
with 

\be
\alpha= &&g^2\frac{E}{\epsilon}\nonumber\\
\left(\frac{\beta}2\right)^2=&& (J+\frac 12)^2+\lambda_{\pm \frac 12}
-g^4\,\,.
\label{D12}
\ee
The equation (\ref{D11}) is a standard hypergeometric equation. The
bound state solutions with $|E|<m$ are Laguerre polynomials for

\be
\alpha= n+\frac 12 +\frac {\beta}2
\label{D13}
\ee
which is the quantization condition with integer $n$ referring 
to the nodal number of the wavefunction as opposed to the radial
quantum number $n_r$ used above. Unwinding (\ref{D13})
in terms of (\ref{D12}) yields the spectra for the relativistic
spin $1/2$ particles/antiparticles

\be
\frac Em\left(1+\frac {g^4}{(n+\frac 12 + \frac
{\beta}2)}\right)^{1/2}=\pm 1
\label{D14}
\ee
with (\ref{D12}) given explicitly by

\be
\frac {\beta}2 = \left|\sqrt{(J+\frac 12)^2 -g^4}\mp \frac 12\right|\,\,,
\label{D15}
\ee
after using (\ref{D10}). This result is in agreement with the original
result established in~\cite{solved_Dirac}. Note that for scalars it agrees
with the semiclassical results we used earlier.

\subsection{Spin 1}

The same analysis performed above can be carried out for the
gauge-independent part of the wavefunction. Indeed, let us 
first consider a {\it massless} gluon in an arbitrary covariant
background field gauge. The equation of motion is standard
and reads

\be
\left(\Box \,\delta_{\mu\nu} -(1-\frac 1{\xi})
\,\nabla_\mu\,\nabla_\nu-2ig\,F_{\mu\nu}\right)\,{\bf a}^\nu=0
\label{G1}
\ee
which simplifies for Feynman gauge $\xi=1$ to 

\be
\left(\Box \,\delta_{\mu\nu} -2ig\,F_{\mu\nu}\right)\,{\bf a}^\nu=0
\label{G2}
\ee
with $\nabla_\mu{\bf a}^\mu=0$. The gluon in (\ref{G2}) has
two physical polarizations, the longitudinal and time-like
ones being gauge artifacts. Generically, we can decompose the 
gluon along its polarizations ${\bf a}^{\mu}={\bf e}^{\mu}_a
{\bf \Psi}^a$, and rewrite (\ref{G2}) in the form

\be
\left(\Box  +\frac g2 \Sigma^{\mu\nu}\,F_{\mu\nu}\right)\,{\bf \Psi}=0
\label{G2x}
\ee
with $i\Sigma_{\mu\nu}/4={\bf e}^T_\mu{\bf e}_\nu$. 
This result is reminiscent of the massless spin $\frac 12$ 
equation (\ref{D1}) if we were to interpret $\Sigma_{\mu\nu}$
as the spin-operator of the gluon in the polarization space
\footnote{The same equation (\ref{G2x}) follows from a path-integral
description of a quantum mechanical evolution of a massless
spin-1 particle in which $\Sigma_{\mu\nu}$ is the covariantized
spin.}.

For {\it massive} gluons the pertinent equation is a variant of
the Proca equation extensively used in the litterature. Here instead,
we proceed by analogy with the spin 1/2 case. In the polarization space 
the equation of motion for spin 1 is just (the g-factor is now 1 instead
of 2)

\be
\left((\Box+m^2) + g\,\vec{\bf S}\cdot (\vec{\bf B}- i\vec{\bf E})\right)
{\bf \Psi} =0\,\,,
\label{G3}
\ee
which is readily checked from the path integral approach
in the first quantization (see also below).
Note that (\ref{G3})  is a $3\times 3$ matrix equation for 
one longitudinal and two transverse polarizations. For spin 1 gluons, 
${\bf S}^{A\,ab}=i\epsilon^{Aab}$. In an external electric
gluon field, (\ref{G3}) reduces to

\be
(\Box+m^2)\,\vec{\bf \Psi} + g\,(\vec{\bf E}\times\,\vec{\bf \Psi})=0\,\,,
\label{G32}
\ee
which shows that the electric field causes the polarization to 
precess in the relativistic equation. To solve (\ref{G3}) in
a Coulomb field ${\bf E}=-g\hat{r}/r^2$ we use exactly the
same method discussed for spin $\frac 12$ except for the use
of vector instead of spinor spherical harmonics,

\be
{\bf \Psi}^a_{JML} = \sum_b
\left(\matrix{J&L&1\cr
-M&(M-b)&b\cr}\right)\,{\bf e}_b^a\,Y_L^{M-b}
\label{G4}
\ee
with $a,b=0,\pm 1$. In this representation the three polarizations are 
chosen real with again ${\bf e}^a_b=\delta^{ab}$. A rerun of the 
precedent arguments show that (\ref{D8}) holds for spin 1 in
a $3\times 3$ matrix form with the substitution

\be
{\bf C} = \left(\matrix{1-(2J+1)&-ig^2\sqrt{\frac{J+1}{2J+1}}&0\cr
-ig^2\sqrt{\frac{J+1}{2J+1}}& 0 &-ig^2\sqrt{\frac{J}{2J+1}}\cr
0&-ig^2\sqrt{\frac{J}{2J+1}}& 1+(2J+1)\cr}\right)\,\,,
\label{G5}
\ee
for $J\neq 0$. The case $J=0$ is special since ${\bf C}={\rm diag}
(0,0,2)$.
The eigenvalues of ${\bf C}$ are solution to a cubic (Cardano)
equation

\be
\lambda^3+2\lambda^2+\lambda(1-(2J+1)^2+g^4)+2g^4=0
\label{G6}\,\,.
\ee
The solutions are all real since the polynomial discrimant
of (\ref{G6}) is negative~\cite{WOLFRAM}. This is expected since the gluon evolution
operator is hermitean. The explicit solutions to (\ref{G6}) are~\cite{WOLFRAM}

\be
\lambda_a=2\sqrt{-Q}\,{\rm cos}\left(\frac{\theta+2\pi\,a}{3}\right)
-\frac 23 
\label{G61}
\ee
with ${\rm cos}\,\theta=R/\sqrt{-Q^3}$ and

\be
Q= &&\frac 13 (1-(2J+1)^2+g^4)-\frac 49\nonumber\\
R=&& \frac 13 (1-(2J+1)^2+g^4)-g^4-\frac 8{27}\,\,\,.
\label{G62}
\ee 

The corresponding spectrum for spin 1 particle is again of the type

\be
\frac Em\left(1+\frac {g^4}{(n+\frac 12 + \frac
{\beta}2)}\right)^{1/2}=\pm 1
\label{G7}
\ee
with 

\be
\left(\frac{\beta}2\right)^2 = (J+\frac 12)^2+\lambda_{0,\pm 1}
-g^4\,\,.
\label{G8}
\ee
The case $J=0$ is special 
and yields $\lambda=(0,0,2)$. 
(Thus one can see that the minimal critical coupling is $\alpha=1/2$, as
for spinless case.) 
Much like the quarks and scalars, the gluons fall onto the
center at a critical coupling which is now set by the branch
point of not only (\ref{G8}) but also (\ref{G61}).

\section{A two-body bound state problem}

\subsection{Generalities}

Non-relativistically, the problem of two bodies (e.g. positronium)
is readily reduced to a single one body problem (hydrogen atom)
by a simple adjustment of the particle mass. Relativistically
this is more subtle.

Let us start with two spinless particles, obeying
two separate Klein Gordon equations. Making 
use of the fact that in the CM the two momenta are equal and opposite
$\vec p=\vec p{(1)}=-\vec p{(2)}$ one can eliminate the relative
energy $p_0{(1)}-p_0{(2)}$ for the total energy
$E=p_0{(1)}+p_0{(2)}$. The resulting relations

\be 
&&\kappa^2-\vec p^2=0\nonumber \\
&&\kappa^2=(E/2)^2-(m_1^2+m_2^2)/2+{(m_1^2-m_2^2)^2 \over 4E^2}
\ee
can be used as the KG equation for two particles with different masses.
The KG simplifies for equal masses, when the last term drops
out.  Fortunately this is is approximately
true for our problem, since for $T$ in the region of interest
the quasiparticle mass difference is relatively small even for $qg$ states.

The quantization follows from the canonical substitution
$E\rightarrow -i\partial_t-V$, $\vec{p}\rightarrow
-i\vec{\partial}-g\vec{A}$. The magnetic effects through
$\vec A$,  are comparable to the electric effects for $v\approx 1$.

\begin{figure}[t]
\centering
\includegraphics[width=3.cm]{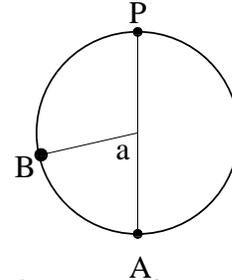}
   \caption{\label{fig_retardation}
Classical motion of two mutually attracting charges.
If the field propagation is instantaneous, a partner of a particle P
is at the ``antipode'' point A. However if particles move
relativistically and the electromagnetic field travels at light speeds,
the particle P sees a field from the earlier position B (drawn for
counterclockwise  rotation).
}
\end{figure}

The next problem is to understand what exactly is the 4 potential
$(V,\vec A)$ in this equation. That is of course the field at 
one particle (say P in Fig.~\ref{fig_retardation}) due to the
other one. Non-relativistically the particle speed is
negligible compared to the speed of light, so one can safely place the
other particle at the opposite point (say A in Fig.~\ref{fig_retardation}).

Including retardation, one classically expects 
the field to emanate from B instead of A. The retarded position B is simply
determined by the condition that the travel time from B to A equals
the time it takes light from B to P, that is $(AB)/v=(PB)/c$. 
In the ultrarelativistic case $v=c$ and one gets a simple
equation for the maximal retardation angle $a$, 

\be 
   a=\sqrt{2+2\,{\rm cos}\,a} 
\ee 
with a root at $a=1.48$. The retardation angle is about
$85^0$ which is rather large. 
    
The classical description  is oversimplified, and in fact
quantum theory allows for field propagation with any speed, not just
c. It demands a convolution of the path (current) with the quantum
propagator of a photon (gluon). At this point, it may appear that
all hope to keep a potential model is lost and a retreat to a full quantum
field theory treatment is inevitable.

This is indeed the case for intermediate coupling $\alpha\approx 1$, but when
it gets stronger one can again claim some virtue in a   potential-based
approach. Our argument presented in~\cite{SZ_CFT} resulted in the
conclusion that the effective photon (gluon) speed is not the usual 
speed of light but {\em larger}, by a parameter ${}^4\sqrt{\lambda}\gg 1$ (where as usual
$\lambda=g^2 N_c$). Thus the field is still dominated by the emission
at the ``antipode'' that is point A instead of  B.  

The magnetic effects are the usual current-current interactions,
present for spinless particles as well, 
plus those induced by (gluo) magnetic moments related with
particle spins, plus their combination (spin-orbit).
The spin effects were argued to be small, see~\cite{BLRS,SZ_CFT}.

For an extensive review
of the known results on how one can reduce a 2-body relativistic
Dirac problem to that of a potential problem we refer to~\cite{HP}.
Here we just note as in~\cite{SZ_CFT} that in the QGP
the quark mass is a ``chiral mass'', so the derivation of the
effective single-body Dirac equation in this case
would be a priori different from the one  discussed in~\cite{HP}.

\subsection{Relativistic Two-Body Bound States}

In this section we illustrate the derivation of the relativistic
two-body bound states between ${\bf qq}$, ${\bf gg}$ and ${\bf qg}$
in QCD by considering the simpler problem in relativistic QED in
both the first and second quantized form. The generalization to
QCD is staightforward in the canonical quantization approach i.e.
$A_0=0$ gauge.

\subsubsection{Second Quantization}

Consider two massive relativistic electrons with Dirac spinors
$\Psi_a$ and $a=1,2$ 

\be
{\cal L}=\sum_{a=1,2} \,\overline{{\bf \Psi}}_a
\,(i\rlap/{\nabla} -{\bf m}_a)\,{\bf \Psi}_a -\frac 14
F_{\mu\nu}F^{\mu\nu}\,\,.
\label{Q1}
\ee
Canonical quantization of (\ref{Q1}) yields the hamiltonian
density

\be
{\cal H} = {\cal H}_F + \frac 12 ({\vec{\bf E}}^2
+{\vec{\bf B}}^2) -\vec{J}\cdot\vec{A} + 
(J_0-\vec{\partial}\cdot{\vec{\bf E}})\,A_0\,\,
\label{Q2}
\ee
with ${\cal H}_F$ the free fermion hamiltonian density and

\be
J^\mu=g\,\sum_{a=1,2} \,\overline{\bf \Psi}_a\gamma^\mu{\bf \Psi}_a\,\,.
\label{Q22}
\ee
The coefficient of $A_0$ (constraint) is Gauss law. The latter
is resolved in terms of the Coulomb potential

\be
\vec{\bf E}_L= -\frac 12 \,J^0\,\partial^{-2}\,J^0\,\,.
\label{Q3}
\ee
Ampere's law follows from the equation of motion for the transverse
part of the electric field

\be
\dot{\vec{\bf E}}_T= -i\left[{\bf H} \,,\,\vec{\bf E}_T\right]
=\partial^2\,{\vec{A}}_T-\vec{J}\,\,.
\label{Q4}
\ee
For stationary (bound) states $\vec{A}_T=(1/\partial^2)\vec{J}$
and the hamiltonian density (\ref{Q2}) simplifies

\be
{\cal H}={\cal H}_F +\frac 12 \vec{\bf E}_L^2
-\frac 12 \vec{\bf B}^2\,\,,
\label{Q5}
\ee
which is the same as

\be
{\cal H}=&&{\bf{\Psi}}^\dagger\left(i\vec{\alpha}\cdot\vec{\partial}\,
+ \beta\, {\bf m}\right){\bf \Psi}\nonumber\\
&&+\frac {g^2}2 {\bf \Psi}^\dagger{\bf \Psi}\frac {-1}{\partial^2}
{\bf \Psi}^\dagger{\bf \Psi}\nonumber\\
&&-\frac {g^2}2 {\bf \Psi}^\dagger
i\alpha^i{\bf \Psi}\frac {-1}{\partial^2}
(\delta^{ij}- \frac {\partial^i\partial^j}{\partial^2})
{\bf \Psi}^\dagger i\alpha^j{\bf \Psi}\,\,,
\label{Q6}
\ee
This form is standard, with the Coulomb (Gauss law) and the 
current-current (Ampere's law) interaction. 
For two particles the spin effects are encoded
in the spinors ${\bf \Psi}$ along with the particle-antiparticle content.
They may be unravelled non-relativistically using a Foldy-Woutuhysen
transformation.

\subsubsection{First Quantization}

Perhaps a more transparent way to address the spin effects
in the presence of gauge fields
when the particle-antiparticle content is not dominant, is
to use the first quantized form of the same problem. For that
it is best to choose the einbein formulation~\cite{CORN} in
the rest frame. For two gauge coupled particles it reads
\footnote{The particle content of the Lagrangian lives only
on the time axis.}

\be
{\cal L}=&& -\frac {V_a}2 (1-\dot{\vec{x}}^2_a) -\frac {{\bf
m}_a^2}{2V_a} +g\,A^0_a-g\dot{\vec x}_a\cdot{\vec A}_a\nonumber\\
&&+\frac 12(\vec{\bf E}^2-\vec{\bf B}^2) + \frac g{V_a}
\vec{\bf S}_a\cdot(\vec{\bf B}_a\mp i\vec{\bf E}_a)
\label{Q7}
\ee
where $a=1,2$ is summed over throughout unless indicated otherwise,
and $A_a=A (x_a)$ and similarly for the fields.
The einbeins are denoted by $V_a$. They will be considered as Lagrange
multipliers at the end and fixed by minimizing the  energy.
The $\mp$ refers to particle-antiparticle.

We note that while the
potentials couple canonically to the currents, the spin couples
canonically to the magnetic field but has an {\it imaginary} coupling
to the electric field in Minkowski space, a situation reminiscent of
Berry phases. Indeed, canonical quantization of (\ref{Q7}) yields
the momenta (only the upper sign is retained from now on)

\be
\vec{p}_a =&&V_a\,\dot{\vec x}_a-g\vec{A}_a\nonumber\\
\vec{\Pi}(x) = &&\vec{\bf E} (x) -
\frac{ig}{V_a}\,\vec{\bf S}_a\,\delta (\vec{x}-\vec{x}_a)\,\,.
\label{Q8}
\ee
The canonical momentum $\vec\Pi$ has a complex shift due to the spin 
of the particle in contrast to the second quantized analysis. By
insisting that (\ref{Q8}) are canonical, we conclude that the 
energy spectrum is shifted by the Berry phase.

The canonical Hamiltonian following from (\ref{Q7}) after
resolving Gauss law reads

\be
{\cal H}= &&\frac 1{2V_a} (\vec{p}^2_a + g\vec{A}_a)^2 
+\frac {{\bf m}_a^2}{2V_a} -\frac {V_a}2 \nonumber\\
&&-g\frac{\vec {\bf S}_a}{V_a}\cdot(\vec{\bf
B}_a-i\vec{\Pi}_{La}-i\vec{\Pi}_{Ta})\nonumber\\
&&+\frac 12 \vec{\Pi}^2_L + \frac 12 (\Pi_T^2+{\vec {\bf B}}^2)^2
-\frac 1{2V_a^2}\vec{\bf S}_a\cdot\vec{\bf S}_a\,\delta(\vec{0})
\label{Q9}
\ee
with

\be
\vec{\Pi}_L (x) =-g\frac
1{\partial^2}\vec{\partial}\,\sum_a\delta(\vec{x}-\vec{x}_a)\,\,,
\label{Q92}
\ee
with the summation over $a$ shown explicitly. 

For stationary states the equation of motions can be used to
solve for $\vec{A}_T$ and therefore for $\vec{\Pi}_T$ and 
$\vec{\bf B}$ through Ampere's and Lenz's law. In particular,

\be
\vec{A}_T (x) = &&-\frac 1{V_a\partial^2} (\vec{p}_a+ g\vec{A}_{Ta})
\,\delta(\vec{x}-\vec{x}_a)\nonumber\\
&&-\frac 1{V_a\partial^2}\,g(\vec{\bf S}_a\times \vec{\partial})\,
\delta(\vec{x}-\vec{x}_a)\,\,.
\label{Q10}
\ee

Noting that the momenta scales with the einbeins as 
$\Pi_L\approx V^0$ and that $\Pi_T\approx {\bf B}\approx 1/V$, it
follows that for stationary states the hamiltonian simplifies to
order ${\cal O} ( 1/{V^2})$, i.e.

\be
{\cal H} \approx  \frac 1{2V_a}\left( \vec{p}^2_a+{\bf m}_a^2 + V^2_a + 2ig
\vec{\bf S}_a\cdot \vec{\Pi}_{La}\right) + \frac 12 {\vec{\Pi}^2_L}\,\,.
\label{Q11}
\ee
The expansion in $1/V^2$ is justified in the non-relativistic limit
since $V\approx {\bf m}$ {\it and} the ultrarelativistic limit since
$V\approx \gamma{\bf m}$ (here $\gamma$ is the Lorentz contraction
factor). Indeed, the extrema 
in $V$ of the total hamiltonian ${\bf H}$ associated to (\ref{Q11}) are
complex and read

\be
V_a=\sqrt{p_a^2+{\bf m}_a^2 + 2ig\,\vec{\bf S}_a\cdot\vec{\Pi}_{La}}
\,\,.
\label{Q12}
\ee
in particular $V\approx \gamma {\bf m}$ as asserted.
For a pair of identical particles ${\bf m}_1={\bf m}_2$ 
in their center of mass frame $\vec{p}_1=-\vec{p}_2=\vec{p}$,
(\ref{Q11}) and (\ref{Q12}) yield 

\be
\frac 14 \left( {\bf H} + \frac {g^2}{x_{12}}\right)^2\approx
\vec{p}^2 +{\bf m}^2 -
2ig^2 (\vec{\bf S}_1-\vec{\bf S}_2)\cdot\vec{\partial}\,\frac 1{x_{12}}
\label{Q13}
\ee
after absorbing the Coulomb self-energy corrections in the
masse ${\bf m}$ . (\ref{Q13}) gives a spectrum analogous to the one described
in the background field section except for the fact that now each of the
two particles carry its own spin in the center of mass frame. Spin-orbit
and spin-spin effects are obtained by carrying the expansion a step
further in $1/V$ in ${\cal H}$.

\section{Bound states in static effective potentials}

Studies of effective potentials in lattice QCD have a long
history. Their $T=0$ version was first obtained in the
classic paper by M.~Creutz who first found confinement on the
lattice in 1979.  The first finite-T results have shown
Debye screening, in agreement with theoretical 
expectations~\cite{Shu_QGP}.  These static
potentials  lead to early conclusions~\cite{MS,KMS} that all
states, even the lowest $\bar c c$ states and $\eta_c,J/\psi$, melt at
$T\approx T_c$. As we already mentioned in the introduction,
these conclusions contradict the recent MEM analysis of the correlators which
indicate that charmonium states stubbornly persist till about $2T_c$.

On theoretical grounds, it has been repeatedly argued (see e.g.
\cite{Shu_how}) that close to $T_c$ the Debye mass is low enough to
allow the color charge to run to rather large values. If so, the binding of
many states must occur, as it was shown in our letter~\cite{IS_newqgp}. 

Recently in a number of publications, the Bielefeld group had
reanalyzed the effective static potentials. One of the key point missed
in previous works is that the expectations values obtained from
thermal
average are by definition the {\em free energy} $F$, related with the
{\em energy} E by the standard thermodynamical relation

\be 
E=F-TS\,\,.\ee
So in order to understand the potential $energy$ one has to remove
the entropy part first. The subtraction results in in much deeper potentials, which
(as we will show below) readily bind heavy (and light) quarks.

\begin{figure}[t]
\centerline{
\includegraphics[width=9.cm]{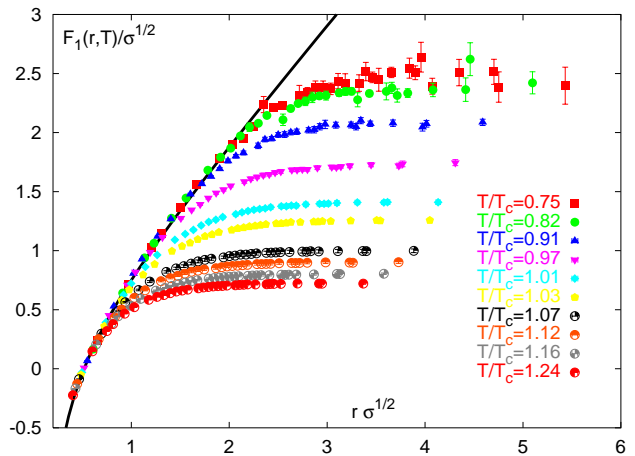}
}
\caption{
Heavy quark free energies in the singlet channel 
for 2-flavors of dynamical quarks at a quark mass of $m/T=0.40$ on $16^3\times 4$
lattices renormalized to the zero-$T$
potential obtained from~\protect\cite{Kaczmarek:2003ph}.
\label{fig_pots}
}
\end{figure}
  A set of potentials obtained by the Bielefeld group is shown in
  Fig.\ref{fig_pots}. The strength of the effective interaction
can be characterize by a combination, called a {\em screening function},
\be S_1(r,T)=-\frac{3}{4} r (F_1(r,T)-F_1(\infty,T))\ee
where the subscript 1 refers to the color singlet  channel
and the 3/4 removes the Casimir for $\bar q q$ representation,
so that $S_1$ is in a way just an effective gauge coupling $\alpha_s$.
A sample of these effective gauge couplings are plotted in Fig.~\ref{fig_screenf}.
The logarithmic plot shows exponential decrease at large r, while the non-logarithmic
plot shows a decrease toward small $r$ due to  asymptotic
freedom. The maximum at $rT\approx 1/2$ indicates that the 
effective Coulomb coupling at $T_c$ is $\alpha_{\rm eff}\approx 4/3 S_1 \approx
1/2$, right in the ballpark used in \cite{BLRS}. One should also note that  
the strength is even larger below $T_c$, where it is related with
confinement at the string breaking point. Finally, we note the surprisingly
monotonous behavior through the phase transition point, due to the
fact that at $T>T_c$ the static charge continues to be screened by a single
light quark, like in a heavy-light (B-like) meson below $T_c$. 

\begin{figure}[t]
\centering
\includegraphics[width=9.cm]{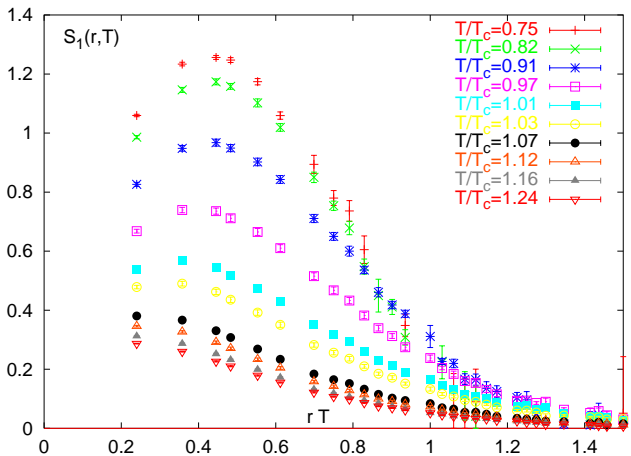}
\caption{
The color singlet screening function, from~\protect\cite{Kaczmarek:2003ph}.}
\label{fig_screenf}
\end{figure}
The effective quark mass, or a constant value of the potential,
was subtracted out: it plays an important role in what follows.

We have parameterized these Bielefeld data by the following expression
(here and below all dimensions are set up by $T_c$, e.g. $T$ means
$T/T_c$ and $r$ is $r T_c$)

\be
F_1(T,r)=1.5-1.1(T-1)^{1/2}-{4\over 3 r}{e^{(-2 T r)}\over {\rm ln\,}(1/r+3T)}
\ee
and then extracted from it the potential energy using 
$E_1=F_1-TdF_1/dT$.

Furthermore, an appropriate normalization of the
potential to be used for the bound state analysis is its decay
at infinity. We use $V(T,r)=E_1(T,r)-E_1(T,r=\infty)$.
(Non-relativistically, the constant part can  be added to a mass.)
The resulting potentials are plotted in Fig.~\ref{fig_my_pots}

Our next next step after extracting the lattice potentials~\ref{fig_my_pots}
is to use them in a  Schrodinger (or appropriate
relativistic) equation  and solve for the bound states. A sample
of the  results obtained using the Klein-Gordon 
equation with this potential is plotted in Fig.~\ref{fig_binding}.
We used a charm quark mass of $1.5\, GeV$ and an effective gluon mass
of $0.6\, GeV$: the results are not very sensitive to it.
\begin{figure}[t]
\centering
\includegraphics[width=7.cm]{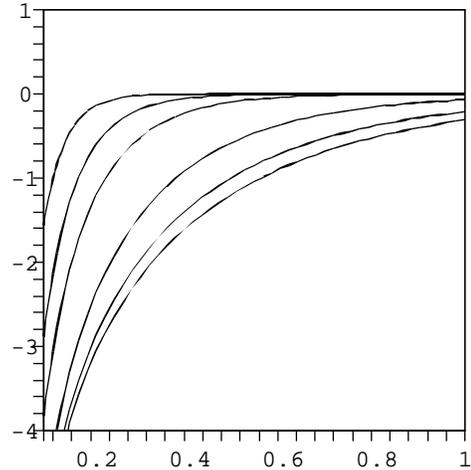}
   \caption{\label{fig_my_pots} 
 The static 
   potentials $V(T,r)$ as a function of the distance $r T_c$.
 The values of temperature used are
   $T=1.,1.2,1.4,2.,
4.,6.,10. T_c$, from right to left.
}
\end{figure}

One can see that charmonium remains bound to about $T=2.7T_c$.  It is
in fact completely consistent with lattice observations~\cite{charmonium} 
using MEM. The fact that the state is traced only
to $T<2T_c$ is completely understandable,
 as for $2<T/T_c<2.7$ it is so weakly bound that the size
of the state may exceed the size of the lattice and could not possible
be seen. Note that at all $T$ the charmonium binding remains rather
small, and so the nonrelativistic treatment of charmonium would be
completely justified. A similar conclusion would be reached for light $\bar q q$ pairs.

This is not the case for the gluonic singlet $gg$ state, which has a color charge larger
by the ratio  $9/4$. We found that the same potential
in this case leads to much larger  binding, reaching  up to 40 percent
of the total mass at $T=T_c$.  There is no question that the relativistic treatment is indeed
needed here. 

We have also looked for $l=1$ states in this potential, which we only found
for the most attractive singlet $gg$ state. Those reache their zero binding at
$T=1.0805\, T_c$. Another next-shell state is an $l=0,n=2$ state, which
exists till  $T<1.205\, T_c$. (Those states are not included in the
calculation of the thermodynamical quantities at the end of the paper.)

\section{Relativistic corrections and  instanton molecules }

\subsection{Relativistic corrections}
\label{sec_rel}

The lattice potentials used above 
were evaluated for {\em static} charges without spin, therefore it does
not include effects proportional to particle velocities and/or spins.
Although we used a relativistic KG equation, we see
that $\bar q q$ (and even the most attractive color singlet $gg$ state)
are not  bound nearly enough as to become massless. On the other hand,
 we know that the lowest $\bar q q$ states, the
pion-sigma multiplet, must do so at $T=T_c$. It means that something is
missing in the interaction. We will discuss those missing effects in
the next section.

The first relativistic effect,
already discussed in  Ref.~\cite{BLRS}, is the 
velocity-velocity force due to magnetic interaction (Ampere's law
in the classical treatment). This corresponds to a substitution of the
effective Coulomb coupling by
\be \alpha\rightarrow \alpha(1-\vec\beta_1 \vec\beta_2)\ee 
where $\beta_i$ are velocities (in units of $c$) of both charges.
In the center of mass the velocities are always opposite, so the
the effective coupling always increases. 

We have estimated the mean velocity squared by using the equation itself
  
\be 
<v^2>=\int dr \chi^2(r) \left(1-\left(\frac
{2M}{E-V(r)}\right)^2\right) 
\ee
where the wave function is appropriately normalized. 
This yields $ <v^2>=.12$ for light quarks at $T/T_c=1.05$.
 If we plug this correction back into the potential, assuming
it scales as $(1+<v^2>)$ as a whole~\footnote{
As we argued above, Ampere's current-current
interaction is $v^2$ times the Coulomb charge-charge interaction, 
but in order for the whole screened static 
potential to scale the same way, other parameters (such as the electric and
magnetic screening lengths) should
coincide, which strictly speaking is not the case.
} we get larger binding (squares in
Fig.~\ref{fig_binding}). With this relativistic correction included,
at $T\approx T_c$ the light quarks become 
relativistic and about as bound as the charmed quarks,
with mean velocity of about 1/3.

  Although we do not yet see how very relativistic motion may come
  about, we will do so in the next section. In anticipation of that,
let us show here what can be a maximal effect of the
the relativistic correction under consideration. If
the particle velocity becomes the speed of light, 
a correction under consideration effectively doubles
the coupling, putting it at $T\approx T_c$ to be $\alpha_{\rm eff}\approx
1$.  If we simply double the effective
potential as a whole, the  binding increases  significantly.
The result is shown by  a single square in Fig.~\ref{fig_binding}(a),
and at $T=T_c$ the binding reaches about 1/4 of the total mass. 
Even larger effect is seen in the particle density at the origin.
As shown in Fig.~\ref{fig_binding}(b)
 it increases by about an order of magnitude.

\begin{figure}[t]
\centering
\includegraphics[width=7.cm]{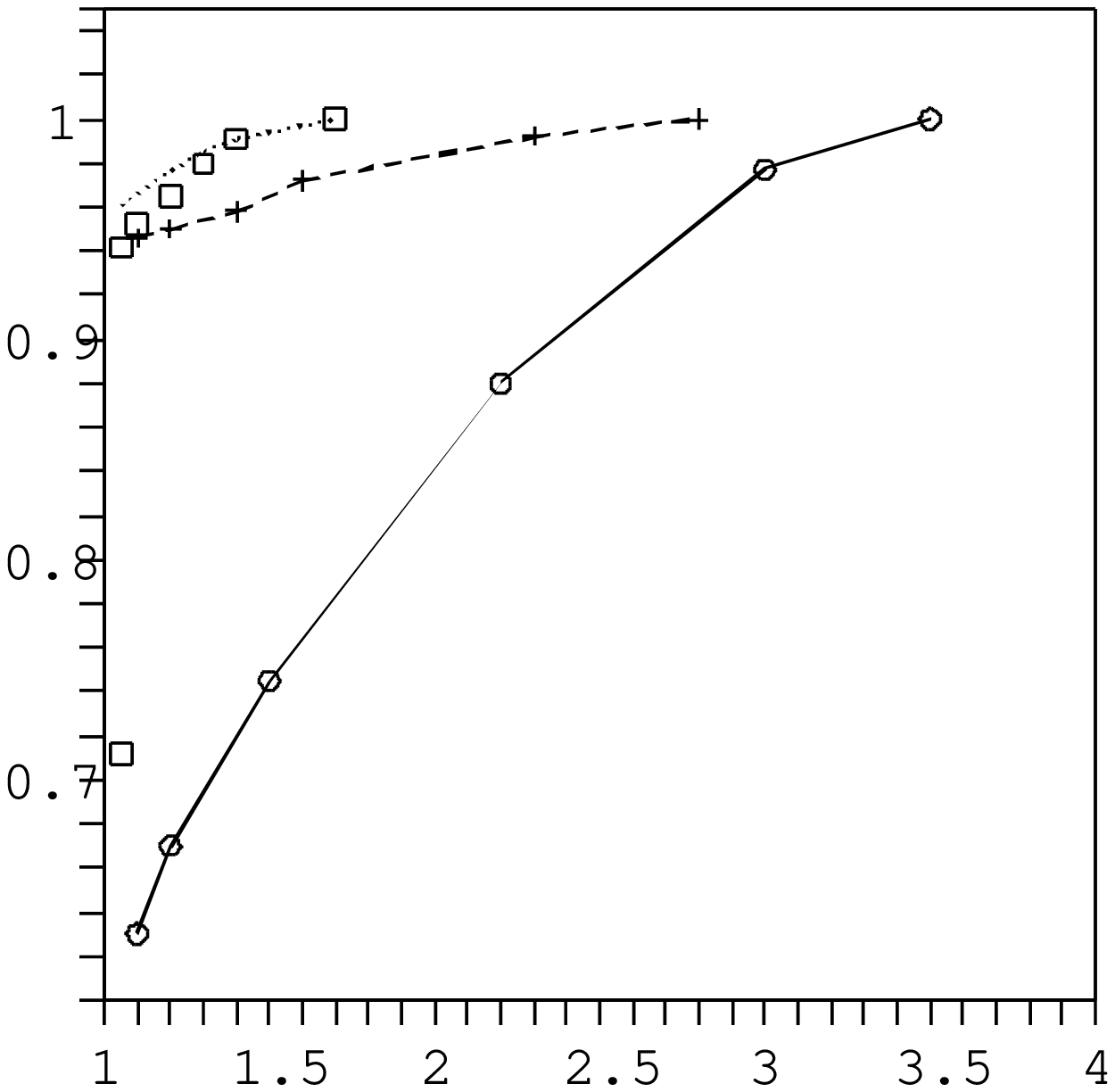}
\includegraphics[width=7.cm]{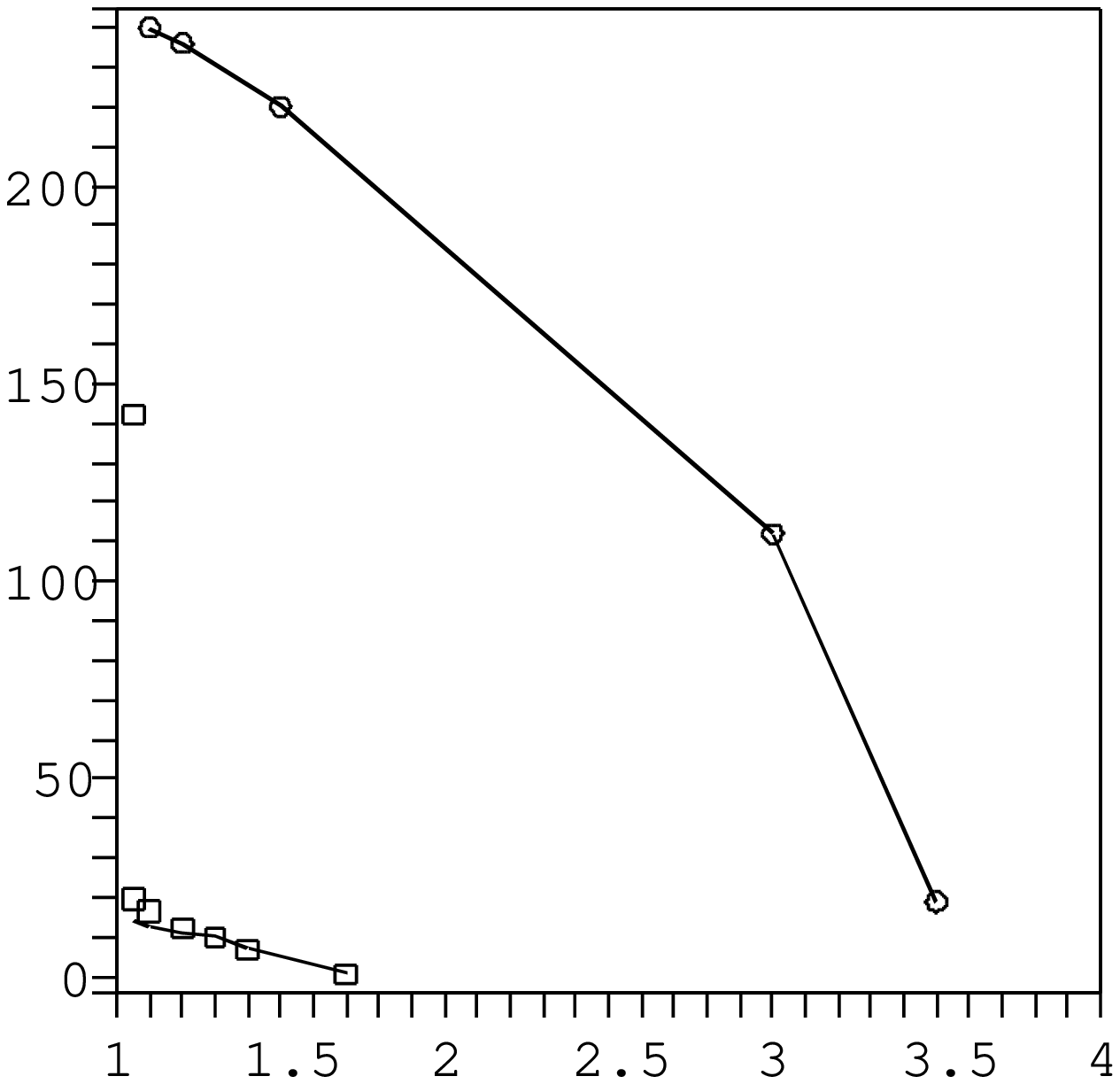}
   \caption{\label{fig_binding}
(a) The energy of the bound state $E$ (in units of the total mass 2M)
versus the temperature $T/T_c$ using 
the lattice effective potential $V(T,r)$,    
for charmonium (crosses and dashed line), singlet light quarks $\bar q
q$ (solid line) and $gg$ (solid line with circles). 
A set of squares show the relativistic
correction to light quark, a
single square at $T=1.05T_c$ is for $\bar q
q$ with twice the coupling, which is the maximal possible relativistic
correction.
(b) The density at the origin $|\psi(0)|^2/T_c^3$ 
 of the bound states 
versus the temperature $T/T_c$
in the lattice effective potential $V(T,r)$,    
for singlet light quarks $\bar q
q$ 
and $gg$
(upper and lower lines with circles). A single square at 
$T=1.05T_c$ is for $\bar q
q$ with twice the coupling. 
}
\end{figure}

\subsection{Interaction induced by the instanton molecules}

 We have seen in the preceding section that relativistic effects
proportional to velocities make all states significantly more
bound and dense at $T\rightarrow T_c$: but
still these effects  are  too weak to bring the
total energy of the lowest $\bar q q$ to zero, as is required
for sigma mesons to trigger a phase transition at $T=T_c$ coming
from above on the temperature axis.

 As discussed  in 
 BLRS paper~\cite{BLRS}, the missing element is 
a quasi-local interaction
induced by the  {\em instanton-antiinstanton molecules}, 
 the lowest clusters of zero total topological charge
allowed in the chirally restored
 phase (see~\cite{SS_98} for a review). 

  In this paper we will not try to estimate
the coupling from first principle but
instead adopt a purely phenomenological approach.
For the local 4-fermion interaction with the coupling constant $G$
the energy shift is given simply by
\be \label{eqn_IAshift}
\delta E= -G |\psi(0)|^2\,\,.\ee
Here we will tune the magnitude of the effective 4-fermion coupling $G_{4q}$
so that the pion-sigma multiplet gets massless
(and then tachyonic) exactly at $T=T_c$.

With $|\delta E|=0.9 2M_q\approx 1.3 {\rm GeV}$ and $|\psi(0)|^2 \sim 350
T_c^3$, after the relativistic correction is included,
 one needs $G\approx 1.5 {\rm GeV}^2$, which is in the
expected ballpark~\footnote{The estimated value is a factor of 2
smaller than used in~\cite{BLRS}, which was derived 
by continuity from NJL-like fits from  $T<T_c$, and somewhat
 overestimated.}
If the relativistic correction would not be there, the value of 
$|\psi(0)|^2$ would be an order of magnitude smaller, see
Fig.~\ref{fig_binding}(b).

The remaining problem is what happens in the glueball ($gg$ singlet)
channel, where we expect the interaction with instantons to be even
stronger than with quarks. Indeed, the instantons are classical
objects
made of gluon field, thus their interaction with gluons would be 
proportional to the action $O(S\approx
8\pi^2 /g^2)$ which is expected to be about 10 times stronger than 
't Hooft's interaction with quarks. If this is the case, the s-wave
glueball state gets tachyonic already at some finite $T$ above $T_c$.

\section{Contribution of the bound states to thermodynamical quantities}

\subsection{The contribution of the quasiparticles}

\setcounter{footnote}{0}

Let us start with the papers by Peshier et al~\cite{Peshier:1995ty} and 
Levai and Heinz~\cite{LH} who used a simple quasiparticle
gas model, deducing what properties the $q,g$ quasiparticles
should have in order to reproduce the  lattice data of the
pressure~\footnote{As all other thermal observables follow from this
function, we will not discuss them.} $p(T)$. Assuming
the usual dispersion relation $\omega^2=p^2+M(T)^2$, one has to deal
with the $T$-dependent of the masses.
For example, for pure glue it is possible to reproduce  $p(T)$ by
assuming $M_g(T)$. 
The qualitative behavior found in~\cite{LH} at high $T$ is  about
linear and rising, as expected perturbatively~\cite{Shu_JETP}. For
$T=(1.3-3)\, T_c$ it is about constant, with a rise towards $T_c$, an
indication of the onset of confinement.

Although these features are qualitatively consistent with the
direct measurements~\cite{Karsch_quasiparticles}, they are not
quantitatively. For $N_f=2$ the expected  masses
at $T=1.5\,T_c$ (close to their minimum) from~\cite{LH} are

\be 
M_g\approx 420 \, {\rm MeV}\,\qquad M_q\approx 300 \, {\rm MeV}\,\,.
\ee
However direct studies by Karsch and Laerman found heavier ones  

\be 
M_g\approx 540 \, {\rm MeV}\,\qquad M_q\approx 620 \, {\rm MeV} \,\,.
\ee
 
If the reader is not impressed by this difference, let us mention that
the corresponding Boltzmann factors for quarks are $exp(-M_q/T)=0.28$
for LH and only $0.075$ for KL values.
This means that the QGP quasiparticles at such $T$ are {\em too heavy} to
reproduce the global thermodynamical observables.

If this numerical example is not convincing, let us go to  
$\cal N$=4 supersymmetric Yang-Mills theory at finite
temperature, for which a parametric statement  can be made.
 At strong coupling
$\lambda=g^2 N_c \gg 1$
the quasiparticle masses are~\cite{Rey_etal} $m\approx\sqrt{\lambda}\, T\gg T$,
and thus the corresponding Boltzmann suppression is about
${\rm exp}(-\#\sqrt{\lambda})$.   In our work~\cite{SZ_CFT} we suggested 
that in this limit the matter is made entirely  of binary bound
states with masses $m\approx T$. We have shown that such light 
and highly relativistic bound  states exist at any coupling,
balanced by high angular momentum $l\approx\sqrt{\lambda}$.
 Furthermore, we have found that the density of such states remains
constant  at arbitrarily large  coupling, although the energy of each
individual state and even its existence depend on $\lambda$.
So,  in this
theory a transition from weak to strong coupling basically implies
a smooth transition from a gas of quasiparticles to a gas of
``dimers''. In QCD going from high $T$ towards $T_c$ 
  the coupling changes  from weak to medium
strong values $\alpha_s=$ 0.5-1 with $\lambda=$ 15-30, and so one naturally
 can see only a half of such phenomenon, with the contribution of the
bound states becoming comparable to that of the quasiparticles.

\subsection{Parameterizations for masses of the bound  states}

As we indicated above,
the pressure problem can be solved
when one accounts for the
additional contribution  of the binary bound states.
We will do so in a slightly schematic way,
combining them into large blocks of states, rather than
going over all attractive channels of Table 1 one at a time.

The ``pion multiplet'' (plus other chiral and $U_A (1)$ partners, e.g.
$\sigma,\eta,\vec\delta$ for $N_f=2$) carry
$2N_f^2$ states, which are to turn massless at $T_c$.
Using (\ref{eqn_IAshift}) for the pion binding and simple
parameterization of the T-dependence of the wave function at the origin
as shown in Fig.~\ref{fig_binding}(b), we arrive at the following
parametrization of their effective mass at temperature $T$

\be 
M_\pi \approx 10\, T_c \left(1-{\rm exp}(-3(T-T_c)/T_c)\right)\,\,,
\ee
which is set to vanish at $T=T_c$. 

For the ``rho multiplet'' (vectors and axial mesons) 
with $6N_f^2$  $\bar q q$ states we use the same expression, but with a
suppression factor of 0.7 in front of the exponent (see~\cite{BLRS} for
details on why instanton molecules are somewhat less effective
in vector channels).

 We have further ignored the (most bound) 
 $gg$ singlets and concentrated on much more numerous
  colored attractive channels.
 Ignoring the differences between
 them for simplicity, we lump them altogether with a mass parameterized as
\be 
M_{\rm colored}\approx 11.5\,T_c\left((T/3T_c)^{0.5}+0.1T_c/(T-T_c)\right) 
\ee
where the second term enforces their disappearance at the critical
point. The corresponding curves for the masses of bound states are shown in
Fig.~\ref{fig_masses}a.
\subsection{The QGP pressure}

In analogy with the so called ``resonance gas'' at $T<T_c$,
the contributions of all  bound states for $T>T_c$ can be simply added
to the statistical sum, as independent particles. However, this is only true
for sufficiently well bound states. The region near the zero binding point
needs special attention. Indeed, in this region 
the bound state becomes virtual and leaves the  physical part of the
complex energy plane: one may naturally expects that its contribution
to the partition function is also going away. 

\setcounter{footnote}{0}

Let us see how this happens, restricting ourselves to the s-wave (l=0)
scattering\footnote{The case of $l>0$ is different, as
the centrifugal barrier allows for quasistationary states and
resonances to exist.}. As it is well known (see e.g. Landau-Lifshits QM, section 131)  
a general amplitude for a system with a shallow level can be written
at small scattering momentum $k$ as

\be 
f(k)= {1 \over -\kappa_0+r_0 k^2/2-ik}
\ee

where $\kappa_0$ is related to the level position and $r_0$ is the so
called interaction range. The subscript refers to 
the $l=0$ partial wave. The cross section is

\be 
\sigma_0={4\pi \over (\kappa_0-r_0 k^2/2)^2+k^2}
\ee
and if the range term is further ignored one gets he familiar
$\sigma=4\pi/m(E+|\epsilon|)$ form which does not care for the sign
of the binding energy $\epsilon=\kappa_0^2/m$.

\begin{figure}[t]
\centering
\includegraphics[width=7.cm]{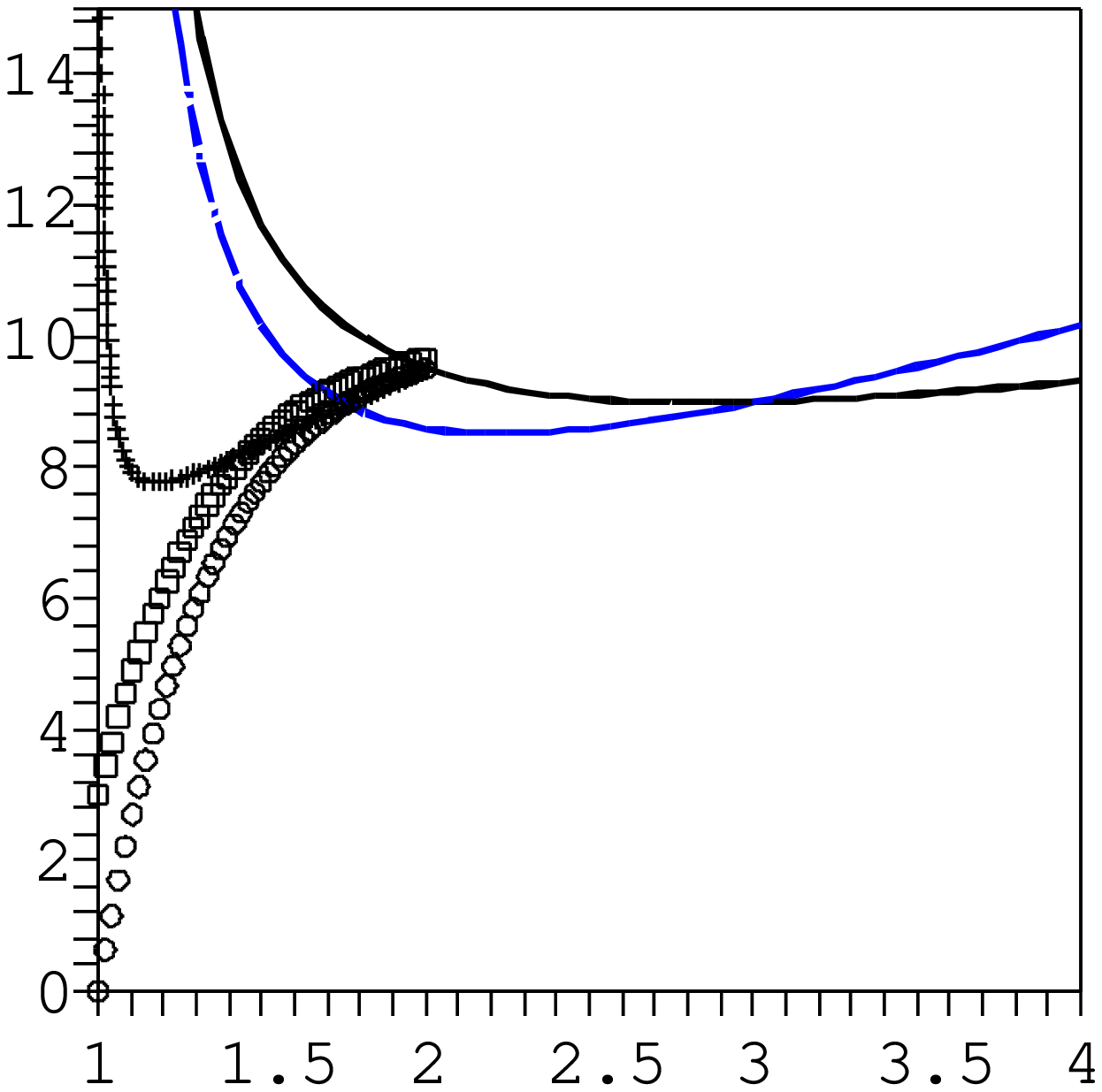}
\includegraphics[width=7.cm]{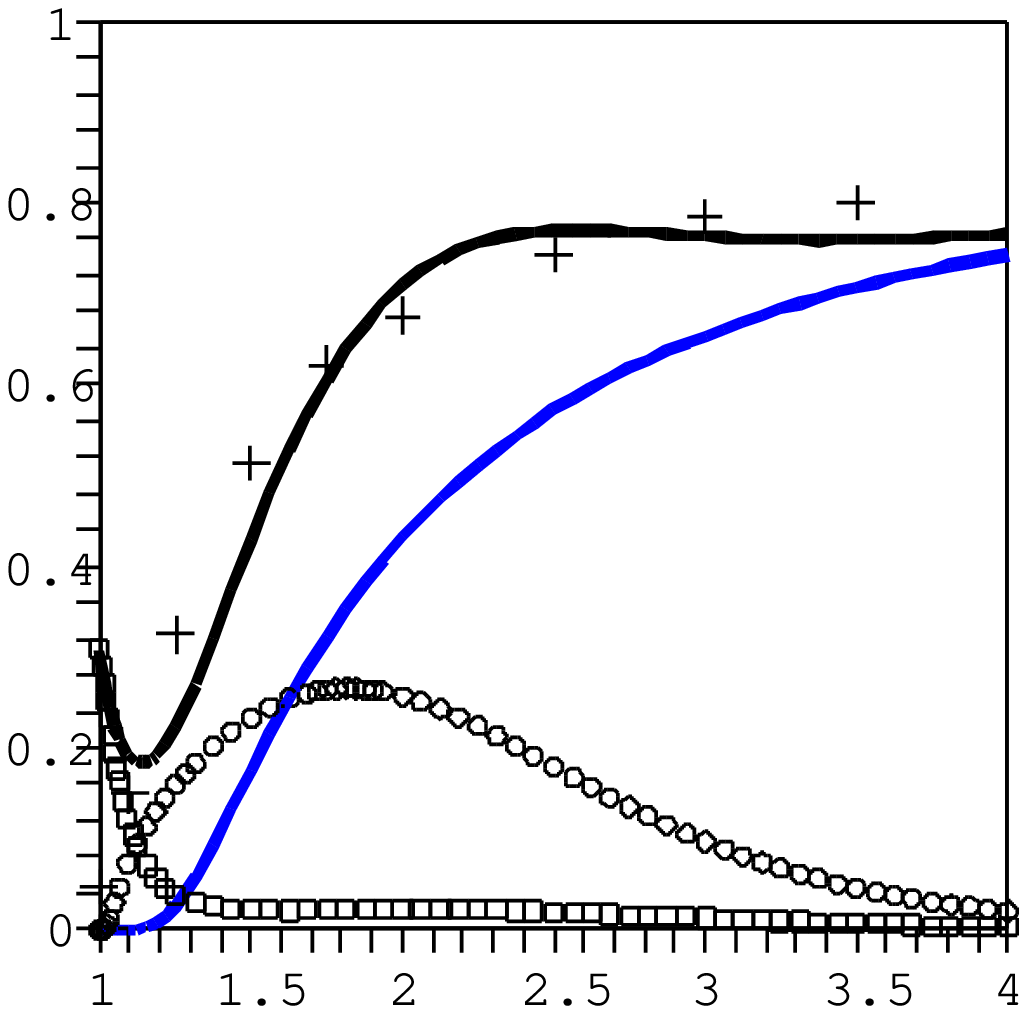}
  \caption{\label{fig_masses}
(a) The lines show twice the effective masses for
quarks and gluons versus temperature $T/T_c$.
Note that for $T<3T_c$ $M_q>M_g$.
Circles and squares indicate estimated masses of the pion-like
and  rho-like $\bar q q$ bound states, while
the crosses stand for 
all colored states.
(b) Pressure (in units of that for a gas of massless and
noninteracting quasiparticles) versus the temperature $T/T_c$.
The crosses  correspond to the $N_f=2$ lattice results, from
Fig.\ref{fig_lat_pressure}. The lower solid curve   
is the contribution of unbound quasiparticles, the upper
includes also that of all bound states.
Squares are for  the pion-like
and  rho-like $\bar q q$ bound states combined,
and circles for all the colored bound
states.}
\end{figure}

 As  $\kappa_0$ goes through zero and changes
sign,
the scattering length $f(k=0)=-1/\kappa_0$ jump from $-\infty$ to $\infty$.
 As one can see from these expressions, a bound level close to zero
generates significant repulsive interaction of the quasiparticles
(with positive energies). As we will see shortly, an account for such
repulsion reduces the contribution of the bound state to the partition
function.

Let us follow the well known Beth-Uhlenbeck expression for non-ideal gases
(see e.g. Landau-Lifshitz, Statistical mechanics, section 77)

\be 
Z_{int}= &&\sum_n e^{-|\epsilon_n|/T}
\nonumber\\&&+{1\over \pi}\sum_l (2l+1)\int_0^\infty
{d\delta_l(k) \over dk} e^{-k^2/mT} dk
\ee
where the first sum runs over all bound states with binding energies
$\epsilon_n$ and the second over the scattering states.
As the simplest example, let us consider the zero binding point $\kappa_0=0$,
for which the expression for the scattering phase can be simplified
to 

\be {\rm exp}(2i\delta_0(k))={r_0 k/2+i \over r_0 k/2-i}     
\ee
and assuming that the temperature $T$ is high enough to ignore the Boltzmann factor
in the integral one gets

\be  
Z_{int}\approx 1+{1\over \pi} (\delta_0(\infty)-\delta_0(0))=1/2\,\,.
\ee
Thus, at the zero binding point the repulsion
reduces the contribution of the bound states
to a half its value. 
As the virtual level moves away from zero, the contribution decreases
further\footnote{Note that this is different from a contribution of a
real resonance, which generates a Breit-Wigner cross section with a 
maximum, and for a narrow resonance
the same contribution as for the bound state.}.  

We will use a simple Fermi-like function to enforce this disappearance
of the level from the statistical sum,
multiplying the level contribution by an additional ``reduction factor''

\be 
R(T)={1\over 1+{\rm exp}[C(T-T_{z.b.})]}\ee
which reduces the contribution at the zero binding point $T_{z.b.}$
by 1/2 and eliminates it at higher $T$. We will use a parameter $C=2/T_c$.

Taking it all together, one finally gets the pressure shown in 
Fig.~\ref{fig_masses}b. One can see that the pions and rhos  peak
at $T_c$\footnote{Recall that the pions gets massless
in the chiral limit. From the current description it looks like
there is a disagreement with the lattice data
at $T=T_c$, but one should recall that the latters are for medium heavy
quarks. 
 }, 
but become unimportant for  higher $T$. The colored states
are too heavy near $T_c$, but then 
their contribution to the pressure becomes comparable to that of the
quasiparticle gas. There are hundreds of them, bringing
quite a large statistical weight, which is however
tamed by a large mass and consequently small Boltzmann factors.  
As a result, the total contribution follows reasonably well the
lattice data points.

\section{Conclusions and discussion}

In this paper we have addressed a number of issues related with the bound
states in the QGP phase at not too high temperature. We have catalogued 
all attractive binary channels, with proper color factors and multiplicities,
see Table 1. We have also presented a unified framework for analysing
two body relativistic bound states of arbitrary spin and mass using
first quantization arguments.

As already mentioned in the Abstract, our main task was to check
the internal
consistency of four (previously disconnected) lattice
results: {\bf i.} spectral densities from MEM analysis of the correlators;
{\bf ii.} static quark free energies $F(R)$; {\bf iii.}  quasiparticle masses; 
{\bf iv.} bulk thermodynamics  $p(T)$. We conclude that everything indeed
fits well together, provided the role of multiple (colored) bound
states is recognised and included.

We have parameterized recent lattice data on free energies for static
quarks, calculated the corresponding effective potentials and solved
the Klein-Gordon equation for charmonium, light quarks and singlet gg
cases. We have found that the bound states exists at $T$ below some
calculated zero-binding points. Our reported range of temperatures
for charmonium and light mesons above $T_c$ agree rather well with
what was seen from lattice correlators using the MEM~\cite{charmonium}. 
 
Our studies of relativistic effects have found that the systems in
question are not yet very relativistic, so that the  relativistic corrections
to the potentials do not exceed 10\%. They have minor effect on
binding, but are more important for the wave function at the origin.
Like the authors in~\cite{BLRS} we concluded that some non-perturbative
interactions for light quarks, completely absent for static ones, should
exist in order to bring the pion and sigma mass  to zero at
$T=T_c$. It is believed to be due to instanton-antiinstanton molecules,
and is quasi-local: thus one can treat this interaction like a
delta-function potential, using the wave function at the origin
calculated without it.

Finally, we have assessed the contribution of all these bound states
to the bulk pressure of the system. We have shown that as the bound
state inches towards its endpoint, its contribution to the pressure 
becomes partially compensated by a repulsive effective interaction
between the unbound quasiparticles. The contribution of the virtual
level above zero quickly disappears\footnote{All of this happens
gradually like in the ionoization process,
so that the disappearence of a state does not lead
to singularities and phase transitions of any kind.}.
 All in all, we have found that 
the summed contributions from the large number of colored bound/free
states above $T_c$ agree well with the bulk lattice pressure.

\vskip .25cm
{\bf Acknowledgments.}
\vskip .2cm

This work was supported in parts by the US-DOE grant DE-FG-88ER40388.
We thank G.~Brown and F.~Karsch 
for many useful discussions, and also  S.~Fortunato,
O.~Kaczmarek and  F.~Zantow who supplied us with
the electronic versions of their unpublished
talks, with useful details on the static lattice potentials.

\vskip .5cm

\section*{Appendix:\\
The Klein-Gordon equation for a Coulomb plus
  quasi-local potential}

In this appendix we will use notations for a single body
KG equation, in schematic notations $(E-V)^2-\vec{p}^2=M^2$,
for a particle of mass $M$ in a potential $V$,
 chosen to be  a superposition of a screened
Coulomb and an additional local term

\be 
V=-{\alpha\over (r+0.001/M)}
\,{\rm exp}(-M_D r)-\tilde U\tilde \delta(r)
\ee
where the $\tilde \delta(r)$ is the ``nonlocal delta function''.
For reasons related with instantons 
we will use it in the form
\be 
\tilde U\tilde \delta(r)=U {1 \over (r^2+\rho^2)^3} 
\ee
with the size parameter $\rho$ to be chosen below to be $\rho=1/(0.6M)$. 
 Note that we have chosen to ignore the 
 coupling constant running, and regulate the Coulomb singularity.

Together with the quasiparticle mass(es) $M$
we thus already have four parameters $\alpha,M_D,U,M$. 
The relevant ones are the  3 dimensionless combinations 
$\alpha,M_D/M,U/M^2$. 
To get familiar with this problem, we  first studied the geometry
of the fixed energy surfaces, $E=const$, in the  parameter
space. A section of those surfaces with two coordinate
planes is shown in Fig.~\ref{fig_binding}.

 In Fig.(a) one can see that
the (regularized)\footnote{
As it is known from the exact solution of the unregularized KG equation,
the solution gets singular at $\alpha=1/2$. Therefore all our results for
 $\alpha>1/2$ are actually sensitive to the regularization used.
} Coulomb and the instanton-induced  potential are kind of
complementary to each other, except near the origin: Coulomb always
have
levels why the quasi-local potential does not for $\tilde U< \tilde
U_{min}\approx 20$.
On another plane, as $M_D$ grows and
 screen the Coulomb field, one needs stronger coupling to keep the
 same binding.

\begin{figure}[t]
\centering
\includegraphics[width=6.cm]{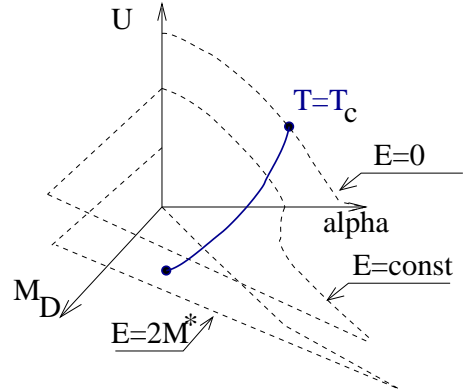}
\includegraphics[width=6.cm]{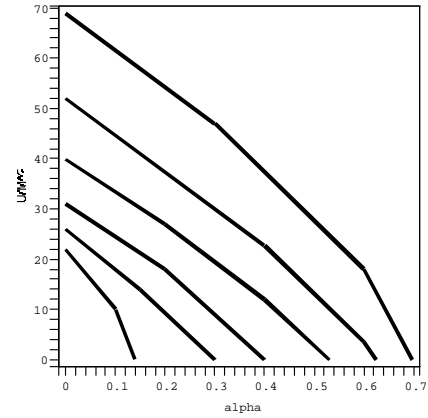}
\includegraphics[width=6.cm]{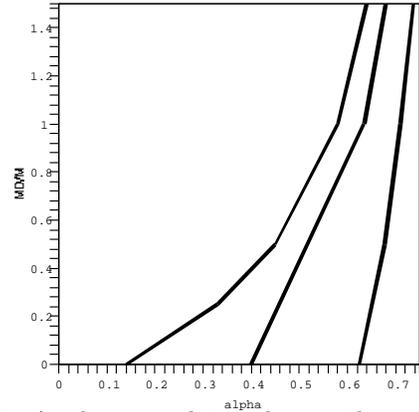}
\caption{\label{fig_app_3d}
A schematic plot explaining the geometry of the surfaces
of fixed binding in the 3-dimensional parameter space.
Its projection  on the $\alpha,U/M^2$ plane 
and on the  $\alpha,M_D/M$ plane  are shown in two subsequent plots.
 In the former 
case  the values of the energy (in units of the mass) is
$E/M=0,0.5,0.75,0.9,0.96,0.99$, from top to bottom. In the
latter case only the lines corresponding
to $E/M=0.5,0.9,0.99$ are shown (right to left).
}
\end{figure}

The very left line, corresponding to $E/M=0.99$ or only 1\% binding,
is close to the ``zero binding line'' (except that it 
reaches the origin $\alpha=0,M_D=0$), to the left of which
the potential in question has {\em no} bound states at all. 

We have not plotted the third projection, as for $\alpha=0$ the value
of $M_D$ is irrelevant and all lines  depend on the $U$ value only.
The zero binding line, separating the unbound states from the bound
ones,  starts at the value of $\tilde
U_{min}\approx 20$ already mentioned. Combining 3 projections for the
same binding, one can now well 
imagine the location and the shape of all constant energy surfaces.

\end{narrowtext}

\begin{thebibliography}{99}
\bibitem{Shu_QGP}
E.~V.~Shuryak, {Phys. Lett.} {\bf B78} (1978) 150,   
{Yadernaya Fizika } {\bf 28}  (1978) 796,
{Phys.Rep.} {\bf 61} (1980) 71. 

\bibitem{LINDE}
A. M. Polyakov, {Phys.Lett.} {\bf B82} (1979) 2410;
A. Linde, Phys. Lett. {\bf 96} (1980) 289.
C. Detar, Phys. Rev. {\bf D32} (1985) 276;
V.~Koch, E.~V.~Shuryak, G.~E.~Brown and A.~D.~Jackson,
Phys.\ Rev.\ {\bf D46} (1992) 3169, 
[Erratum-ibid.\ {\bf D47} (1993) 2157]
T.H. Hansson and I. Zahed, Nucl. Phys. {\bf B374} (1992) 277;
T.H. Hansson, M. Sporre and I. Zahed, Nucl. Phys. {\bf B427} (1994) 447; 
T.H. Hansson, J. Wirstam and I. Zahed, Phys. Rev. {\bf D58} (1998) 065012; 

\bibitem{IS_newqgp}E.~V.~Shuryak and I.~Zahed,
{\tt hep-ph/0307267}, submitted to PRL.


\bibitem{hydro}  
D. Teaney, J. Lauret and E.V.~Shuryak, 
{ Phys. Rev. Lett. }{\bf 86} (2001) 4783, more details in nucl-th/0110037.
P.F. Kolb, P.Huovinen, U. Heinz, H. Heiselberg,
{ Phys. Lett.} {\bf B500} (2001)  232.
More details in P.~F.~Kolb and U.~Heinz,
{\tt nucl-th/0305084}.


\bibitem{GM}D.~Molnar and M.~Gyulassy,
Nucl.\ Phys.\ {\bf A697} (2002)  495;
[Erratum-ibid.\ {\bf A703} (2002) 893]


\bibitem{Teaney_visc} 
D.~Teaney {\tt nucl-th/0301099}

\bibitem{ATOM1}
J.Cubizolles et al., {\tt cond-mat/0308018}; K.Strecker et
  al., {\tt cond-mat/0308318}

\bibitem{ATOM2}
M.W. Zwierlein et al.,  {\tt cond-mat/0311617}.



\bibitem{ADSCFT1} 
J.M.~Maldacena, 
Adv.~Theor.~Math.~Phys. {\bf 2} (1998) 231, {\tt hep-th/9711200}.



\bibitem{ADSCFT2}
G.T.~Horowitz and A.~Strominger, { Nucl. Phys.} {\bf B360} (1991) 197.
S.S.~Gubser, I.R.~Klebanov and A.A.~Tseytlin, {Nucl.\ Phys.\ } {\bf B534} (1998) 202.



\bibitem{atoms_flow} 
K.M.~O'Hara et al,
  Science {\bf 298} (2002) 2179;
T.~Bourdel et al, Phys. Rev. Lett.  {\bf 91} (2003)  020402.

\bibitem{PSS}
G.~Policastro, D.~T.~Son and A.~O.~Starinets,
Phys.\ Rev.\ Lett.\  {\bf 87} (2001) 081601.


\bibitem{MS}
T.~Matsui and H.~Satz,
{Phys.\ Lett.\ } {\bf B178} (1986)  416.

\bibitem{KMS}
F.~Karsch, M.~T.~Mehr and H.~Satz,
Z.\ Phys.\ {\bf C37} (1988)  617.

\bibitem{charmonium}
S.~Datta, F.~Karsch, P.~Petreczky and I.~Wetzorke,
{\tt hep-lat/0208012};
M. Asakawa and T. Hatsuda,
Nucl. Phys. {\bf A715} (2003) 863c; 
{\tt hep-lat/0308034}

\bibitem{Karsch:2002wv}
F.~Karsch et al
Nucl.\ Phys.\  {\bf B715} (2003)  701.


\bibitem{Karsch_quasiparticles}
P. Petreczky, F. Karsch, E. Laermann,
S. Stickan, I. Wetzorke,  Nucl. Phys. Proc. Suppl. {\bf 106} (2002) 513.

\bibitem{Karsch_pressure}
F.~Karsch,
AIP Conf.\ Proc.\  {\bf 631}  (2003) 112.


\bibitem{SZ_CFT} 
E.V. Shuryak and I. Zahed, 
Phys.\ Rev.\  {\bf D69} (2004) 014011.



 \bibitem{Shu_JETP} 
E.V.~Shuryak, Zh.E.T.F {\bf 74} (1978) 408,
   (Sov. Phys. JETP {\bf 47} (1978) 212), 


\bibitem{solved_Dirac} 
C.G.~Darwin, Proc. Roy. Soc. Lond. Ser. {\bf A118} (1928)  654,
W.~Gordon, Z. Physik {\bf 48} (1928) 11.


\bibitem{BLRS} G.~E.~Brown, C.~H.~Lee, M.~Rho and E.~Shuryak,
{\tt hep-ph/0312175}.


\bibitem{Edmonds}
A.~R.~Edmonds, ``Angular Momentum in Quantum Mechanics'', Ed. Princeton 
(1974), pp 83-85.


\bibitem{WOLFRAM}
S.~Wolfram,~{\it http://mathworld.wolfram.com/CubicEquation.html}.

\bibitem{HP} 
V.~Hund and H.~Pilkuhn, J. Phys. {\bf B33} (2000) 1617


\bibitem{CORN}
A. Kihlberg, R. Cornelius and M. Mukunda, Phys. Rev. {\bf D23} (1981) 2201.


\bibitem{Shu_how}
E.~V.~Shuryak,
Nucl.\ Phys.\ {\bf A717} (2003) 291.

\bibitem{Kaczmarek:2003ph}
O.~Kaczmarek, S.~Ejiri, F.~Karsch, E.~Laermann and F.~Zantow,
{\tt hep-lat/0312015}.

\bibitem{SS_98} 
T.~Schafer and E.~V.~Shuryak,
Rev.\ Mod.\ Phys.\  {\bf 70} (1998) 323


\bibitem{Peshier:1995ty}
A.~Peshier, B.~Kampfer, O.~P.~Pavlenko and G.~Soff,
Phys.\ Rev.\ {\bf D54} (1996) 2399.


 \bibitem{LH}
P.~Levai and U.~W.~Heinz,
Phys.\ Rev.\  {\bf C57} (1998)  1879.

\bibitem{Rey_etal}
S.J.~Rey, S.~Theisen and J.T.~Yee, 
Nucl. Phys. {\bf B527} (1998) 171.









\end{thebibliography}
\end{document}